\documentclass[prl,twocolumn,amsmath]{revtex4}
\usepackage{graphicx}
\usepackage{dcolumn}
\usepackage{bm}
\usepackage[usenames]{color}

\begin{document}

\title{Ultra-high sensitivity magnetic field and magnetization measurements with an atomic magnetometer}
\author{H. B. Dang$^1$ and A.C. Maloof$^2$ and  M. V. Romalis$^1$}
\affiliation{$^1$Department of Physics,  Princeton University,
Princeton, New Jersey 08544, USA\\
$^2$ Department of Geosciences,  Princeton University, Princeton,
New Jersey 08544, USA}

\begin{abstract}
We describe an ultra-sensitive atomic magnetometer using
optically-pumped potassium atoms operating in spin-exchange
relaxation free (SERF) regime. We demonstrate magnetic field
sensitivity of 160 aT/Hz$^{1/2}$ in a gradiometer arrangement with a
measurement volume of 0.45 cm$^3$ and energy resolution per unit
time of $44 \hbar$. As an example of a new application enabled by
such a magnetometer we describe measurements of weak remnant rock
magnetization as a function of temperature with a sensitivity on the
order of 10$^{-10}$ emu/cm$^3$/Hz$^{1/2}$ and temperatures up to
420$^\circ$C.
\end{abstract}
 \maketitle



High sensitivity magnetometery is used in many fields of science,
including physics, biology, neuroscience, materials science and
geology. Traditionally low-temperature SQUID magnetometers have been
used for most demanding applications, but recent development of
atomic magnetometers with sub-femtotesla sensitivity has opened new
possibilities for ultra-sensitive magnetometery \cite{MagNatRev}.

Here we report new results of sensitive magnetic field measurements
using a spin-exchange relaxation-free potassium magnetometer. By
eliminating several sources of ambient magnetic field noise and
optimizing operation of the magnetometer we achieve magnetic field
sensitivity of 160 aT/Hz$^{1/2}$ at 40 Hz. The measurement volume
used to obtain this sensitivity is 0.45 cm$^3$, resulting in a
magnetic field energy resolution of $VB^2/2\mu_0 = 44 \hbar$,  a
factor of 10 smaller than previously achieved with atomic
magnetometers \cite{Kominis}. Energy resolution on the order of
$\hbar$ has been realized with SQUIDs at high frequency and
milli-Kelvin temperatures with small input coils \cite{Awschalom,
Clarke}. However for cm-sized SQUID sensors operating at 4.2 K the
energy resolution at low frequency is typically several hundreds
$\hbar$ \cite{Drung,Carelli,Seppa} and the magnetic field
sensitivity is about 1 fT/Hz$^{1/2}$ \cite{Drung1}.

When comparing various magnetometery techniques it is important to
distinguish between applications requiring detection of smallest
magnetic moments and those requiring detection of smallest
magnetizations. For the former, it is usually advantageous to use
the smallest possible sensor. For example, magnetic resonance force
microscopy (MRFM) can detect a single electron spin \cite{Rugar}. On
the other hand, for detection of very weak magnetization one needs a
sensor with the highest magnetic field sensitivity, since $B \sim
\mu_0 M$ in the vicinity of the source. For example, recently
developed magnetometers using a single nitrogen-vacancy (NV) center
in diamond are promising for detection of single electron and
nuclear spins because of their small size \cite{Maze,Jelezko}. In
diamond crystals with larger concentration of NV centers the
magnetic field sensitivity is limited by dipolar interactions with
other inactive color centers and has been optimistically projected
at $10^{-16}$ T/Hz$^{1/2}$/cm$^{3/2}$ \cite{Taylor}, which is the
level already realized experimentally in this work.

One of the well-developed magnetometry applications requiring high
magnetization sensitivity is paleomagnetism \cite{paleo}. Analysis
of magnitude and direction of remnant magnetization in ancient rocks
provides  geological information going back billions of years and
has been used, for example, to establish the latitudinal
distribution of continents through time and to provide a critical
test of the theory of plate tectonics. The magnetization is usually
carried by low concentrations of tiny igneous crystals or
sedimentary grains of magnetite or hematite, leading to very weak
bulk magnetization. SQUID-based rock magnetometers are widely used
for studies of such samples.

Here we demonstrate measurements of weakly-magnetized rock samples
with higher sensitivity than possible with SQUID-based magnetometers
\cite{Kirschvink}. Equally important, the measurements are performed
continuously  as a function of temperature with temperatures up to
420$^{\circ}$C. Such temperature-dependent studies are crucial for
paleomagnetic measurements as it allows one to separate the
contribution of recently acquired magnetization and understand which
magnetic minerals are present in the sample. In the past such
measurements could only be obtained by repeated heating and cooling
cycles of the sample or with much lower sensitivity using a variable
temperature vibrating sample magnetometer \cite{Gallet}. Atomic
magnetometry thus allows more sensitive magnetization measurements
over a wider range of temperatures than is possible with any other
detector.

The basic principles of spin-exchange relaxation free alkali-metal
magnetometers have been described in \cite{Allred,Savukov}. They are
based on the observation that the dominant source of spin relaxation
due to spin-exchange collision in alkali-metal vapor is suppressed
at high alkali-metal density in a low magnetic field \cite{Happer}.
The fundamental sensitivity limits of such magnetometers due to spin
projection noise are estimated to be on the order of 10$^{-17}$
fT/Hz$^{1/2}$/cm$^{3/2}$. The main experimental challenge in
achieving such sensitivity is to minimize ambient sources of
magnetic field noise.

Superconducting  shields can in principle provide a magnetic
noise-free environment. However, in applications requiring
measurements on samples above cryogenic temperature one is often
limited by ``dewar noise''- magnetic field noise on the order of
several fT/Hz$^{1/2}$ generated by conducting radiation shields used
for thermal insulation \cite{Nenonen}. For this work we use a
ferrite magnetic shield, first introduced in \cite{Kornack}, as the
inner-most shield layer. Low electrical conductivity of ferrite
materials eliminates magnetic noise generated by Johnson currents
allowing one to reach magnetic noise level below 1 fT/Hz$^{1/2}$.
Other conducting materials in the vicinity of the magnetic sensor
can also generate Johnson magnetic noise. We recently developed a
new method based on fluctuations-dissipation theorem to estimate
magnetic noise from conductors in various geometries \cite{SKLee}.
It allows us to identity the largest sources of noise and select
appropriate experimental components.

A drawing of the magnetometer apparatus is shown in Fig. 1. The
sample, such as a weakly magnetized rock, is introduced through a
12~mm ID quartz tube that passes through the apparatus. The sample
can be heated in-situ by electric heaters with AC current at 20 kHz.
To reduce magnetic noise from the sample heating wires, they are
removed from the vicinity of the magnetometer and heat is
transmitted by diamond strips held with AlN cement. To thermally
isolate the sample heater from the magnetometer a radiation shield
is constructed by depositing a thin film of gold through a fine wire
mesh on a glass slide. A spherical glass cell 23 mm in diameter
containing K metal, 60 torr of N$_2$ and 3 atm of $^4$He gas is
heated to 200$^{^\circ}$C in a boron-nitride oven with AC electric
heaters. Magnetic fields and first-order gradients are controlled by
a set of coils wound on a G-7 fiberglass frame. The same frame also
incorporates water cooling. The apparatus is enclosed in the ferrite
magnetic shield with a diameter and length of 10 cm and operates in
a vacuum of 1 mTorr. Not shown in the figure are a 316 stainless
steel vacuum vessel and two additional layers of mu-metal magnetic
shields.

\begin{figure}[h]
\includegraphics[width=3.5in] {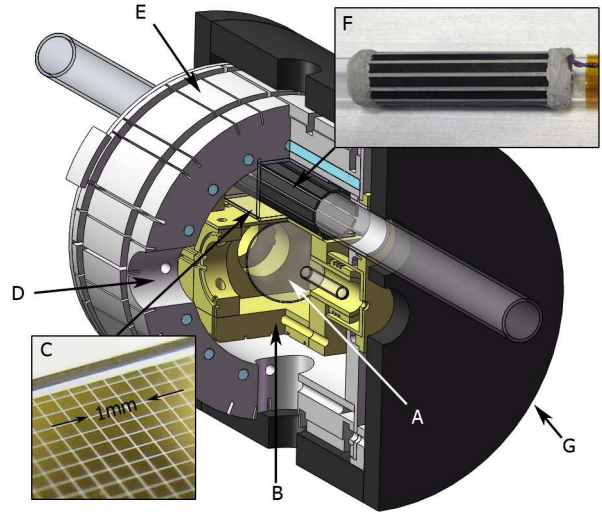}
\caption{Drawing of the magnetometer system. A- alkali-metal cell,
B-boron-nitride oven, C- photo of the radiation shield, D- optical
access for probe laser, E- G-7 fiberglass frame for magnetic field
and gradient coils and water cooling, F - photo of the sample
heater, G- ferrite magnetic shield. }\centering
\end{figure}

The magnetometer is operated using 773 nm DFB lasers which are
cooled to -20$^{^\circ}$C to reach K D1 line at 770
nm\textcolor{blue}{;}  no additional wavelength or intensity
feedback is needed. The pump laser is tuned near the resonance and
directed along the $\hat{z}$ axis, while the probe laser is detuned
by about 0.5 nm to the red side of the resonance and is sent along
the $\hat{x}$ axis.  The polarization of the probe laser is measured
using Faraday modulation technique. All three components of the
magnetic field are zeroed and field gradients are adjusted to
maximize the signal. Gradiometric measurements are performed by
imaging the probe beam onto a two-channel photodiode and taking the
difference between the two channels. The effective distance between
the two channels, called the baseline of the gradiometer, is equal
to 0.5 cm, determined by applying a calibrated magnetic field
gradient. The distance between the sample and the sensor volume is
equal to 2.4 cm, determined from the ratio of the signals in the two
channels of the gradiometer. The sensing volume for each channel is
$0.5\times 0.5\times 1.8 $ cm$^3$. Atom diffusion plays a minor role
on the time scale of spin relaxation.

To understand the performance of the magnetometer we have developed
a detailed model of the optical rotation signal that incorporates
various relaxation effects for the K vapor and the absorption of the
pump laser as it propagates into the optically-dense cell
\cite{Walker}. A plot of the optical rotation signal in response to
a 10 pT low-frequency excitation field is shown in Fig. 2, comparing
the results of the model and experimental measurements. The
alkali-metal density and the resonance linewidth are measured
independently, so there are no free parameters in the model. The
overall size of rotation is in good agreement with predictions. The
probe polarization rotation noise is about $1.5\times 10^{-8}$
rad/Hz$^{1/2}$, limited by photon shot noise. So the sensitivity of
the magnetometer can reach $5 \times 10^{-17}$ T/Hz$^{1/2}$ under
optical conditions in the absence of any environmental magnetic
noise.

\begin{figure}[h]
\includegraphics[width=3.5in] {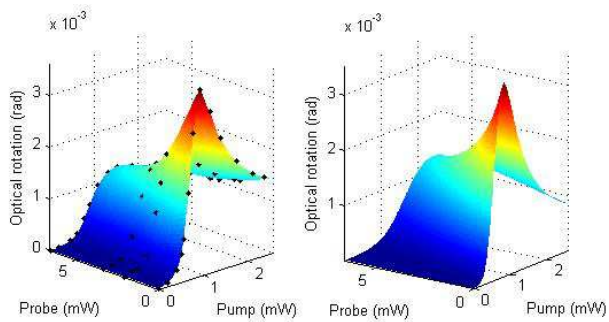}
\caption{Optical rotation in response to a 10 pT oscillating
magnetic field as a function of the pump and probe laser power. Left
panel - measurement points with color interpolation, right panel-
theoretical calculation. }\centering
\end{figure}

In practice\textcolor{blue}{,} to realize the lowest measured
magnetic noise \textcolor{blue}{,} we need to somewhat compromise
the performance of the magnetometer to operate in the regime of
lowest environmental magnetic noise. While the sensitivity of the
magnetometer is largest for magnetic fields with a frequency below
its natural bandwidth of about 3 Hz, the magnetic field noise
increases at low frequency. This $1/f$ noise is due to hysteresis
losses in the ferrite shield and is a feature of all magnetic
materials \cite{Kornack}. The ferrite material used for the shield
was chosen for its low loss factor. We introduce a bias $B_z$ field
to shift the magnetometer resonance and the peak of the magnetic
field response to higher frequencies. Unfortunately, this technique
reduces the magnetometer signal since only the co-rotating component
of an oscillating magnetic field excites the spins.

In Fig. 3 we summarize the magnetic noise measurements. The data are
recorded for several values of the bias field $B_z$ and the magnetic
field response is shown in the top panel. The bottom panel shows the
magnetic field noise from a single magnetometer channel and the
noise obtained from a gradient measurement. The noise in the
difference of the two channels is divided by $\sqrt{2}$ to determine
the intrinsic sensitivity of each channel. The optical rotation
noise, recorded in the absence of the pump beam, is also shown. The
intrinsic magnetic field noise obtained from the gradiometer
measurements reaches 160 aT/Hz$^{1/2}$ at 40 Hz. It is still not
limited by optical rotation noise and is probably due to imperfect
cancellation of the ferrite noise or local sources of magnetic
fields, such as produced by droplets of K metal. The magnetization
sensitivity of the gradiometer for our geometry with a 1 cm$^3$
sample reaches $6\times 10^{-11}$~emu/cm$^3$/Hz$^{1/2}$.

\begin{figure}[h]
\includegraphics[width=3.5in] {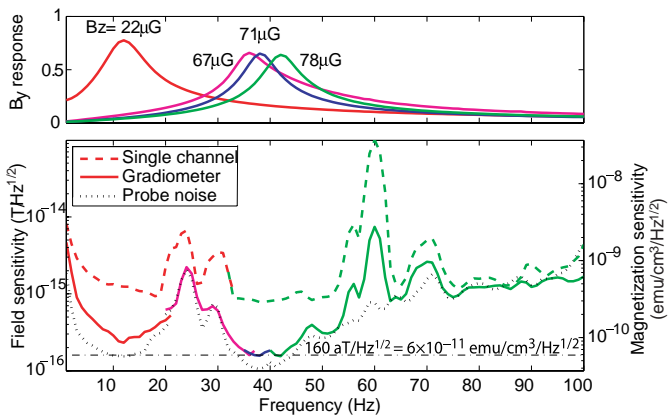}
\caption{Magnetic field response curves (top) and noise spectrum
(bottom) for different values of the $B_z$ magnetic field. The left
axis shows magnetization sensitivity of the gradiometer. }\centering
\end{figure}

The material samples are introduced into the apparatus through the
access tube at ambient pressure. The  sample is held at the end of
high purity quartz tube by pumping on its other end with a vacuum
pump. We found that all glues that we tested have a large magnetic
contamination and most of them would not survive heating to required
temperatures, while the quartz tube did not present significant
background after thorough cleaning. The quartz tube and the sample
are rotated around the axis at about 7 Hz to distinguish sample
magnetic fields from constant backgrounds and move the signal to a
region of lower magnetic field noise.

A 9 mm diameter, 13 mm long cylinder was prepared from a sample of
very weakly magnetized $\sim$635 Ma Ravensthroat Formation peloidal
Dolostone from the Mackenzie Mountains, Canada. Two vector
components of the rock magnetization were determined  by measuring
the phase of the recorded signal relative to the sample rotation
phase. The absolute value of the magnetization transverse to the
rotation axis is plotted in Figure 4 as a function of temperature.
The sample is continuously rotated and slowly heated over a period
of about 2 hours. The magnetization drop at 300-350$^{\circ}$C is
due to unblocking of pyrrhotite or titanomagnetite crystals with the
remaining magnetization most likely carried by magnetite.
Measurements of all 3 vector components of the magnetization can
allow one to obtain the direction of  primary sedimentary
magnetization.

\begin{figure}[h]
\includegraphics[width=3.5in] {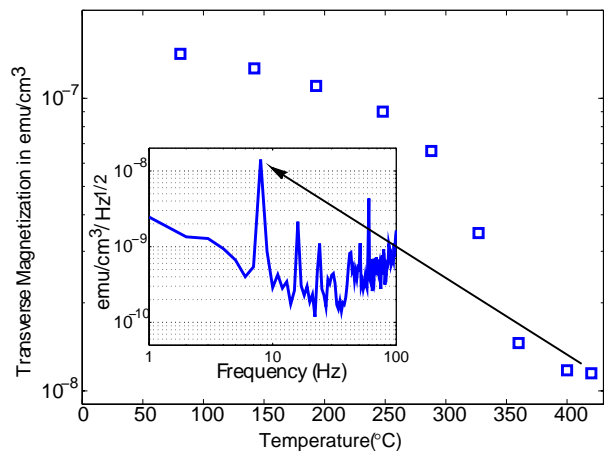}
\caption{Rock magnetization as a function of temperature. Inset
shows that even at highest temperature the signal-to-noise is
greater than 30. These measurements are obtained without
gradiometric recording. }\centering
\end{figure}

In summary, we have demonstrated what is to our knowledge the most
sensitive cm-sized detector of magnetic fields and magnetization
operating at low frequency. The absence of cryogenics allows for
much larger thermal power dissipation, so sample temperatures can be
varied over a wide range without extensive radiation shielding.  We
have achieved sample temperatures up to 500${^\circ}$C and higher
temperatures should be possible with thicker heating wires. Samples
have also been maintained at room temperature by gently blowing air
through the sample tube.  The small size, low laser power and
continuous magnetic field recording allow versatile use of the
magnetometer. In addition to the paleomagnetic application explored
here, many other uses can be readily implemented, including
detection of magnetic nanoparticles \cite{Chemla}, NMR
\cite{Ledbetter} and weak high-temperature ferromagnetic ordering
\cite{Young}. The fundamental sensitivity limits of SERF
magnetometers have not yet been reached, so further improvements can
be expected. We would like to thank John Tarduno for useful
discussions. This work was supported by NSF grant PHY-0653433.


\begin{thebibliography}{99}
\bibitem{MagNatRev}  D. Budker and M. Romalis, Nature Physics \textbf{3},
227 (2007).
\bibitem{Kominis} I.K. Kominis, T.W. Kornack, J.C. Allred, and M.V. Romalis,  Nature {\bf 422}, 596
(2003).
\bibitem{Awschalom} D.D. Awschalom {\it et al.} Appl. Phys. Lett. {\bf 53}, 2108 (1988).
\bibitem{Clarke} M. M\"{u}ck, J. B. Kycia, and J. Clarke, Appl.
Phys. Lett. {\bf 78} 967 (2001).
\bibitem{Drung} D. Drung {\it et al}, IEEE Trans. Appl. Supercond. {\bf 17}, 699 (2007).
\bibitem{Carelli} P. Carelli, M. G. Castellano, Physica B {\bf 280},
537 (2000).
\bibitem{Seppa} P. Hakonen,M. Kiviranta, H. Sepp\"{a}, J. Low
Temp. Phys. {\bf 135}, 823 (2004).
\bibitem{Drung1} D.Drung, Physica C {\bf 368}, 134  (2002).
\bibitem{Rugar} D. Rugar, R. Budakian, H. J. Mamin  and B. W. Chui, Nature {\bf 430}, 329
(2004).
\bibitem{Maze} J. R. Maze {\it et al.}, Nature {\bf 455}, 644
(2008).
\bibitem{Jelezko} G. Balasubramanian {\it et al.}, Nature {\bf 455}, 648
(2008).
\bibitem{Taylor} J. M. Taylor {\it et al.}, Nature Physics {\bf 4}, 810
(2008).
\bibitem{paleo} {\it Paleomagnetic Principles and Practice}, Lisa Tauxe,
Kluwer (2002).
\bibitem{Kirschvink} J. L. Kirschvink, R. E. Kopp, T. D. Raub, C. T. Baumgartner and J. W. Holt,
Geochem. Geophys. Geosys. {\bf 9}, Q05Y01 (2008).
\bibitem{Gallet} M. Le Goff and Y. Gallet, Earth and Planet. Sci. Lett.
{\bf 229}, 31 (2004).
\bibitem{Allred} J. C. Allred, R. N. Lyman, T. W. Kornack and M. V.
Romalis, Phys. Rev. Lett. {\bf 89}, 130801 (2002).
\bibitem{Savukov} I. M. Savukov and M. V. Romalis, Phys. Rev. A {\bf 71} 023405
(2005).
\bibitem{Happer} W. Happer and H. Tang, Phys. Rev. Lett. {\bf 31}, 273
(1973).
\bibitem{Nenonen} J. Nenonen, J. Montonen, and T. Katila, Rev. Sci. Instrum. {\bf 67}, 2397
(1996).
\bibitem{Kornack} T. W. Kornack, S. J. Smullin, S.-K. Lee, and M. V. Romalis, Appl.
Phys. Lett. {\bf 90}, 223501 (2007).
\bibitem{SKLee} S.-K. Lee and M. V. Romalis, J. Appl. Phys. {\bf 103}, 084904
(2008).
\bibitem{Walker} T. G. Walker and W. Happer, Rev. Mod. Phys. {\bf 69}, 629
(1997).
 \bibitem{Chemla} Chemla Y.R. {\it et al}, PNAS  {\bf 97}, 14268 (2000).
\bibitem{Ledbetter} Ledbetter M.P. {\it et al}  PNAS {\bf 105}, 2286-2290 (2008).
\bibitem{Young} Young D.P. {\it et al}  Nature {\bf 397}, 412 (1999).



\end{thebibliography}
\end{document}